\documentclass[prl,aps,twocolumn,showpacs]{revtex4}
\usepackage{amssymb}
\usepackage{amsmath}
\usepackage{graphicx}

\begin{document}
\title{3rd Order Temporal Correlation Function of Pseudo-Thermal Light}
\author{Yu Zhou, Jianbin Liu, and Yanhua Shih}
\affiliation{Department of Physics, University of Maryland, Baltimore
County, Baltimore, MD 21250}

\begin{abstract}
This experiment reports a nontrivial third-order temporal
correlation of chaotic-thermal light in which the randomly radiated
thermal light is observed to have a 6-times greater chance of being
captured by three individual photodetectors simultaneously than that
of being captured by three photodetectors at different times
(separated by the coherent time of pseudo-thermal light), indicating
a ``three-photon bunching" effect. The nontrivial correlation of
thermal light is the result of multi-photon interference.
\end{abstract}
\maketitle

In 1956,  Hanbruy-Brown and Twiss (HBT) discovered a surprising
photon bunching effect  of thermal light\cite{HBT}.  In a spatial
HBT interferometer, the spatially, randomly distributed thermal
light was observed to have a two-times greater chance of being
captured by two individual photodetectors in a small, transverse
area of correlation equal to the coherent area of the thermal
radiation than that of being captured in different coherent areas.
For a large angular sized thermal source the spatial correlation is
effectively within a physical ``point". The point-to-point
near-field spatial correlation of thermal light has been utilized
for reproducing nonlocal ghost images in a lensless configuration
\cite{Gpaper}. In a temporal HBT interferometer, the temporally,
randomly radiated photons seem to have a two-times greater chance to
be measured within a short time window which equals the coherence
time of the thermal field than that of being measured in two
different coherence time windows. For a natural thermal radiation
source, such as the Sun, its coherence time could be as short as
femtoseconds. It seems that the photons are more likely to come
together and arrive at the detectors simultaneously if we try to
detect them at some special space-time positions. This behavior is
phenomenologically explained as ``photon bunching" \cite{bunching}.
What is the physical cause of photon bunching? We believe this is a
reasonable question to ask. It seems that there is no reason to have
photons created in pairs from a thermal source. The radiation
process of thermal light is stochastic.  The radiated photons should
be created randomly in the source, rather than bunching in pairs. In
fact, this bunching effect seems even more strange for $N$th-order
correlation measurements of thermal light, in which more than two
photo-detections are involved. Quantum theory predicts that $N$
photons have $N!$ times greater chance of being captured within the
same coherent time window of the light field than that of being
captured in $N$ different coherent time window of the light
field\cite{Liu}.

Recently, several papers reported high order ghost imaging and ghost
interference experiments by using a pseudo-thermal
source\cite{highorder}. In their experiments, the authors treated
the electromagnetic field from the pseudo-thermal source as Gaussian
random variables. Since for Gaussian random variables any moments of
order higher than 2 can always be expressed in terms of the first
and second order moments, they measured high order moments of
pseudo-thermal light by measuring the first and
 second order moments with one or two CCD's, then calculated the higher order moments\cite{goodman}.

In this letter, we wish to report a three-photon temporal
correlation measurement of pseudo-thermal light in which three
photodetectors are involved in a three-fold joint measurement. The
difference between our experiment and the experiments mentioned
above is that in our experiment we use single photon detectors to
directly measure the high order correlation function of
pseudo-thermal light. We use quantum theory to describe the light
field because it reveals the underlying physics and is applicable to
the case when the intensity is so low that only a few photons are in
the system. The experimental results show that the three individual
photodetectors $D_1$, $D_2$ and $D_3$ have a six times greater
chance of being triggered at $t_1 = t_2 = t_3$ than that of being
triggered at $t_1 \neq t_2 \neq t_3$, apparently indicating that the
photons are emitted in ``triples" by the thermal source.

\begin{figure}[hbt]
\centering
\includegraphics[width=75mm]{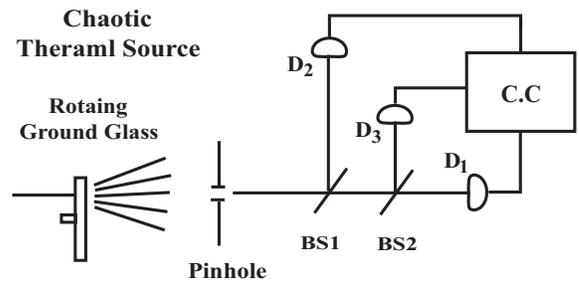}
\caption{Schematic setup of the experiment. }
\label{scheme}
\end{figure}
The schematic experimental setup is shown in Fig.~\ref{scheme}. The
light source is a standard pseudo-thermal source, which contains a
CW laser beam, a fast rotating diffusing ground glass and a focal
lens (with $25.4mm$ focal length). The $632.8$nm laser beam is
focused by a lens, onto the rotating ground glass, to a diameter of
$\leq $100$\mu$m. The coherent laser beam is scattered by the fast
rotating ground glass to simulate a thermal field with $\sim$$0.2
\mu s$ coherence time. The coherence time  of pseudo-thermal light
is determined by the angular speed of the disk, the curvature of the
focused laser beam and the transverse distance between the pinhole
and the laser beam, the detailed discussion can be found in
Ref.\cite{correlationtime}. Effectively, the rotating ground glass
produces a large number of independent point sub-sources (3 to 5
$\mu$m in diameter) with independent random phases.  A pinhole with
a diameter size $\sim 1$mm is placed $800$mm from the ground glass
to select a small portion of the radiation within its spatial
coherence area. At this distance the size of coherent area is about
$\sim 10$mm. The pseudo-thermal light passes through beam-splitters
BS$1$ and BS$2$. The transmitted radiation is detected by
photodetector $D_1$, the reflected radiations are detected by
photodetectors $D_2$ and $D_3$, respectively. To simplify the
discussion, we achieved equal intensities in the three paths by
manipulating the transimission-reflection coefficients of the two
beamsplitters and had equal distances between the light source and
the three photodetectors.  $D_1$, $D_2$, and $D_3$ are fast
avalanche photodiodes working in the photon counting regime. The
photo-detection response time is on the order of a few hundred
picoseconds, which is much shorter than the $\sim 0.2 \mu s$
coherence time of the radiation.  The output pulses from $D_1$,
$D_2$ and $D_3$ are sent to a three-fold coincidence counting
circuit which provides a three-photon counting histogram as a
function of $t_1 - t_2$ and $t_1 - t_3$, where $t_j$, $j =1,2,3$, is
the registration time of the photo-detection event at $D_1$, $D_2$
and $D_3$, respectively.

\begin{figure}
\centering
\includegraphics[width=1.6in]{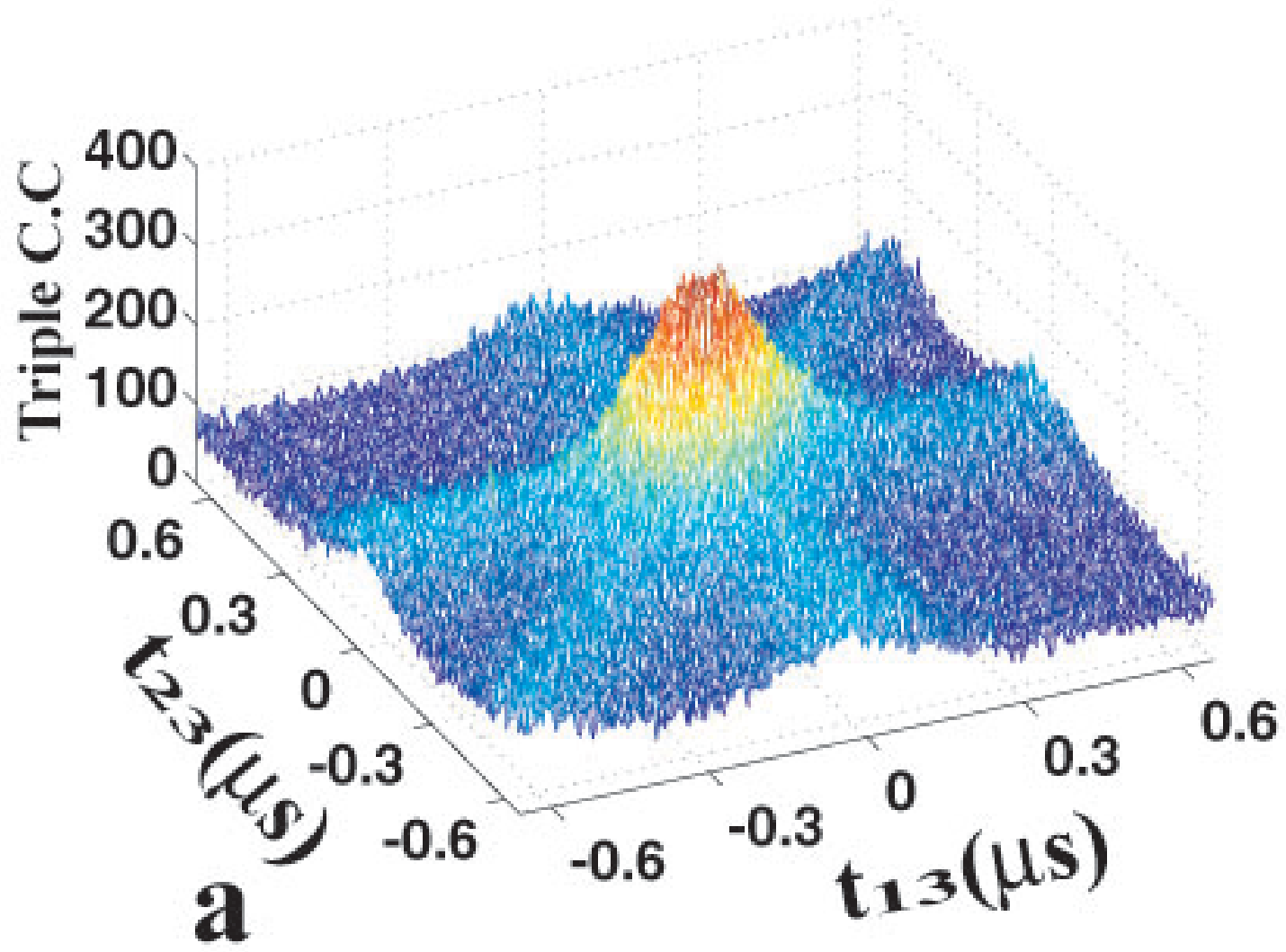}%
\hspace{1mm}%
\includegraphics[width=1.6in]{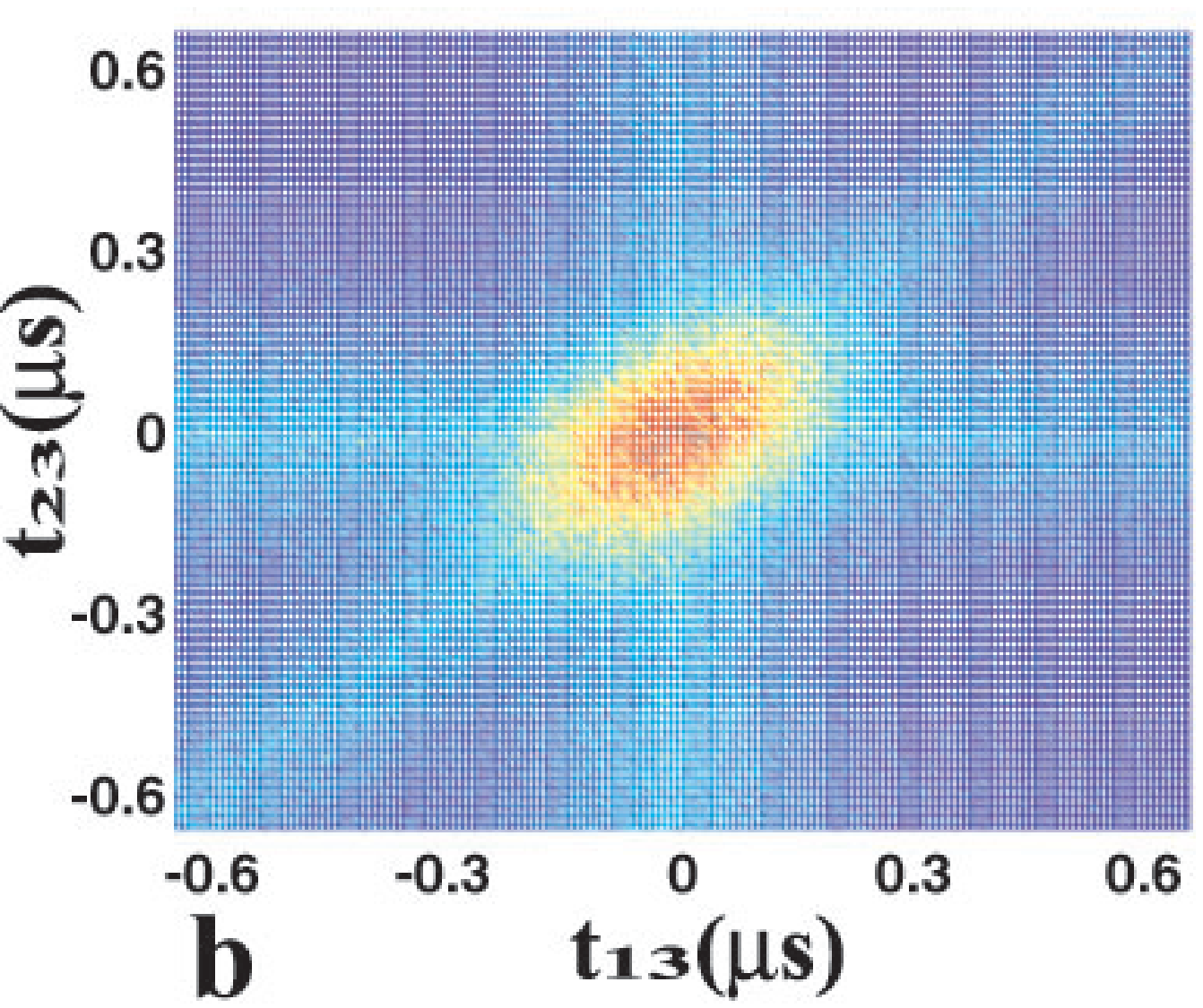}\\
\includegraphics[width=1.6in]{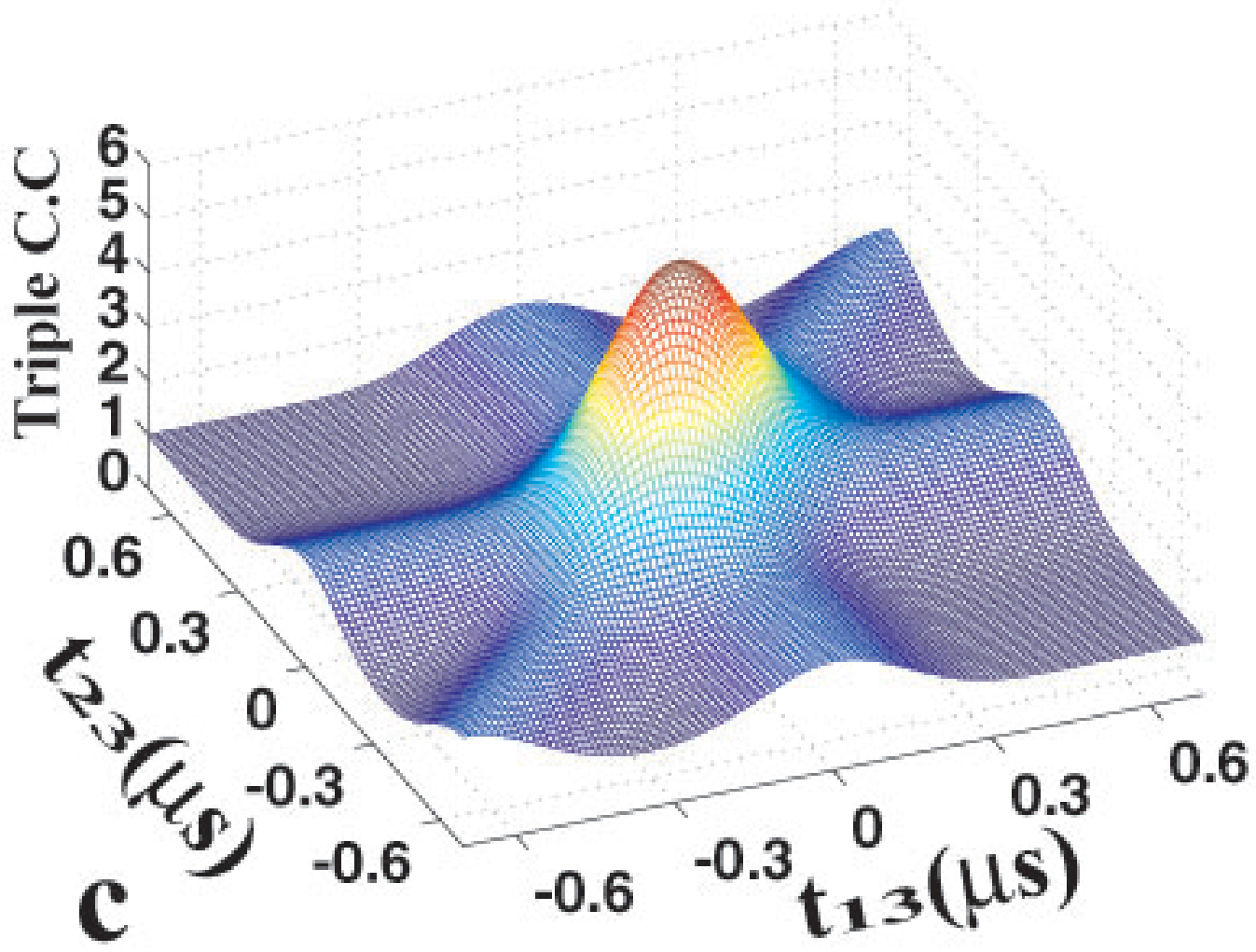}%
\hspace{1mm}%
\includegraphics[width=1.6in]{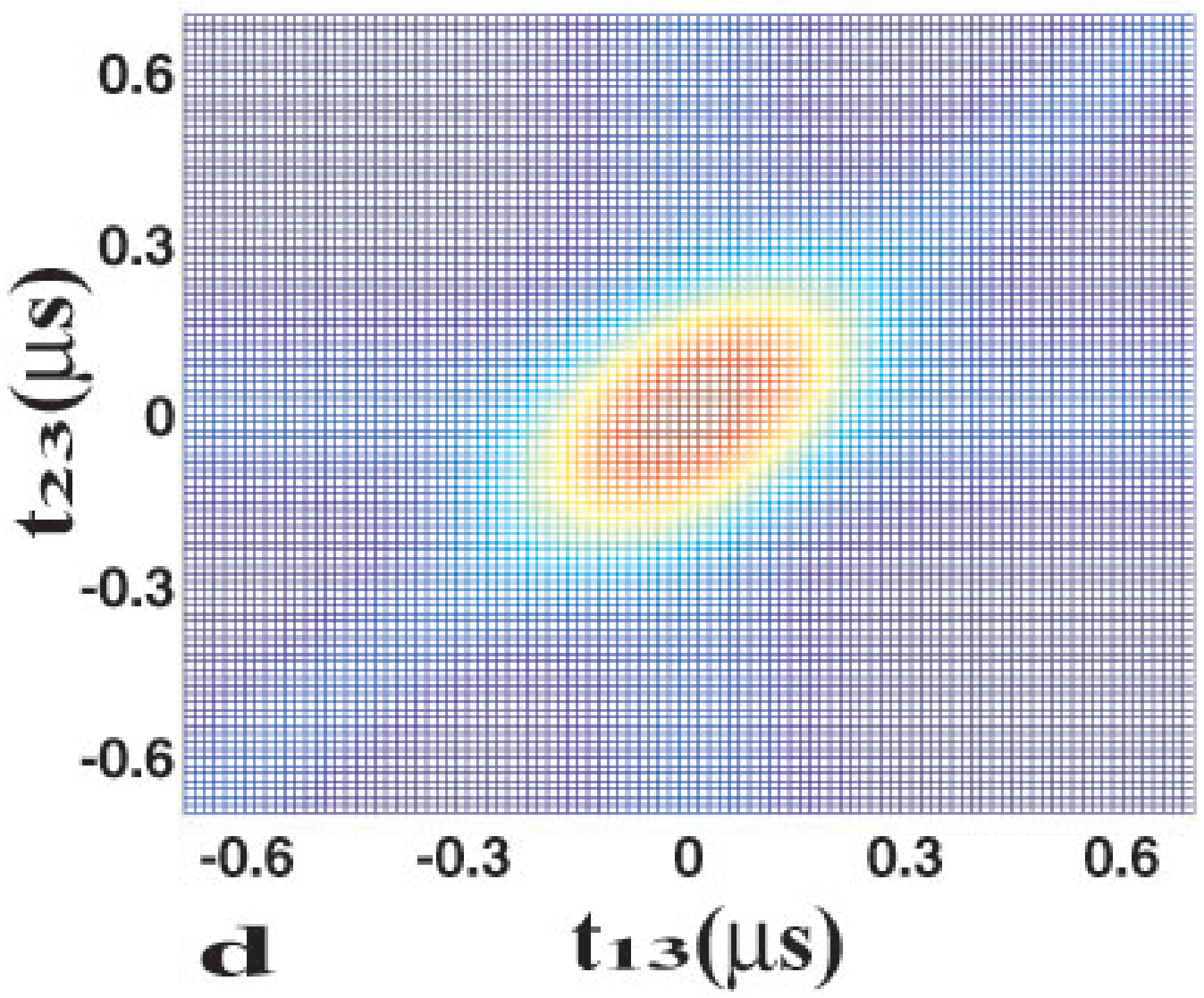}
\caption{(Color Online) Measured (upper, a and b) and calculated
(lower, c and d) third-order temporal correlation function of
thermal light. The 3-D three-photon joint detection histogram is
plotted as a function of $t_{13} \equiv t_1 - t_3$ and $t_{23}
\equiv t_2 - t_3$. The simulation function is calculated from
Eq.~(\ref{g3-1}). In addition, the single detector counting rates
for $D_1$, $D_2$, and $D_3$ are all monitored to be constants.}
\label{simuexp}
\end{figure}

The experimentally measured  and  simulated 3-D third-order temporal
correlation functions are reported in the upper and lower parts of
Fig.~\ref{simuexp}, respectively.  The simulation is calculated
based on Eq.~(\ref{g3-1}).  It is easy to see that (1) the measured
correlation function is close to the simulated function, and (2) the
randomly radiated ``thermal photons" have much greater chance of
being jointly detected in triples when $t_1 = t_2 = t_3$ than that
of being detected when $t_1 \neq t_2 \neq t_3$.

\begin{figure}[hbt]
\centering
\includegraphics[width=70mm]{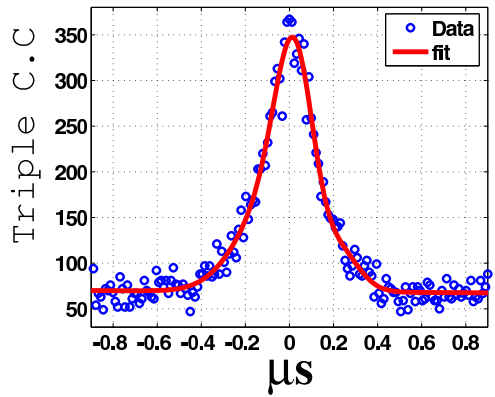}
\caption{(Color Online) Crosse section of the three-photon
coincidence counting histogram, which is sliced from the top left
corner to the bottom right corner of Fig.~\ref{simuexp}b. The
contrast between the maximum counting rate and the constant
background is $4.9\pm 0.25$ to 1, indicating a nontrivial
third-order correlation function with visibility of $\sim66\%$.}
\label{crossection}
\end{figure}

From Eq.~(\ref{g3-1}) we know that in order to claim the measured
correlation is a third order effect the contrast between the peak
and the background should be lager than $4$ to $1$, corresponding to
a visibility larger than $60\%$. To compare the joint counting rate
at $t_1 = t_2 = t_3$ (three-photon bunching) with the joint counting
rate at $t_1 \neq t_2 \neq t_3$, a sliced ``cross section" of the
measured 3-D histogram is illustrated in Fig.~\ref{crossection}. The
plot is a 2-D cross section of Fig.~\ref{simuexp}(b) sliced from the
top left corner to the bottom right corner.  The contrast between
the maximum counting rate, which occurs at $t_1 = t_2 = t_3$, and
the constant background is $4.9\pm 0.25$ to 1, indicating a
nontrivial third-order correlation function with visibility of
$\sim66\%$, which is greater than the 2 to 1 contrast ($33\%$
visibility) of HBT.   This result shows that thermal light has a
much greater chance to be bunched in triples, rather than in pairs.
The theoretically expected contrast is 6 to 1 ($71\%$ visibility).
In addition, the single detector counting rates for $D_1$, $D_2$,
and $D_3$ are all monitored to be constant. The reason for observing
a lower visibility is due to the finite size of the detector.  We
expect to achieve higher visibility by simulating ideal point
detectors.

Phenomenologically, we may name the nontrivial three-photon
correlation ``three-photon bunching" because in the measurement we
find that the photons do have $3!$-times greater chance of being
jointly captured by $D_1$, $D_2$, and $D_3$ simultaneously than that
of being jointly captured at different time(separated by coherent
time of the light field). However, does this observation really mean
that three photons have $3!$-times greater chance to be emitted as a
bunch from the thermal source? As we have discussed earlier, we have
difficulties following the photon bunching argument when $N$ takes a
large value.

In the view of quantum interference, the observed three-photon
correlation of thermal radiation is the result of \emph{three-photon
interference}, which involves the
nonlocal superposition of three-photon amplitudes, a nonclassical
entity corresponding to different yet indistinguishable alternative
ways of triggering a three-photon joint-detection event.
The probability of observing a three-photon joint-detection event
is calculated from the Glauber theory \cite{Glauber}
\begin{align}\label{QM-2nd-00}
& \ \ \ \ G^{(3)}(t_1, t_2, t_3 ) \\ \nonumber
&= \big{\langle}  \hat{E}^{(-)}(t_1)\hat{E}^{(-)}(t_2)\hat{E}^{(-)}(t_3)
\hat{E}^{(+)}(t_3)\hat{E}^{(+)}(t_2)\hat{E}^{(+)}(t_1) \big{\rangle}
\end{align}
where $\langle ... \rangle$ denotes an expectation value operation based on the
quantum state of the measured electromagnetic field.
Thermal radiation is typically in mixed states.  The generalized
density operator for a chaotic-thermal field can be written as
\cite{klyshko}
\begin{equation}\label{ThermalState}
\hat{\rho}=\sum_{\{n\}} \, p_{\{n\}} \, |\{n\} \rangle \langle \{n\}|,
\end{equation}
where $p_{\{n\}}$ is the probability that the thermal field is in
the state
$$
|\{n\} \rangle \equiv \prod_{\omega} |n_{\omega} \rangle
= |n_{\omega} \rangle
|n_{\omega'} \rangle ...
|n_{\omega^{\prime \prime ... \prime}} \rangle.
$$
Here we have simplified the problem to 1-D (temporal) with one
polarization. The summation of Eq.~(\ref{ThermalState}) includes
all possible frequency modes $\omega$, occupation numbers
$n_{\omega}$ for the mode $\omega$ and all possible combinations
of occupation numbers for different modes in a set of $\{n\}$.
The quantized field operator takes the following form
\begin{equation}
\hat{E}^{(-)}_j= \int d\omega \, f(\omega)\, g(\omega, t_j, z_j) \, \hat{a}^{\dag} ({\omega}),
\end{equation}
where $f(\omega)$ is the spectral distribution function of the field,
$g(\omega, t_j, z_j)$ is the Green's function which propagates each mode
of the field from the source to the $j$th detector.

\begin{figure}[hbt]
\centering
\includegraphics[width=80mm]{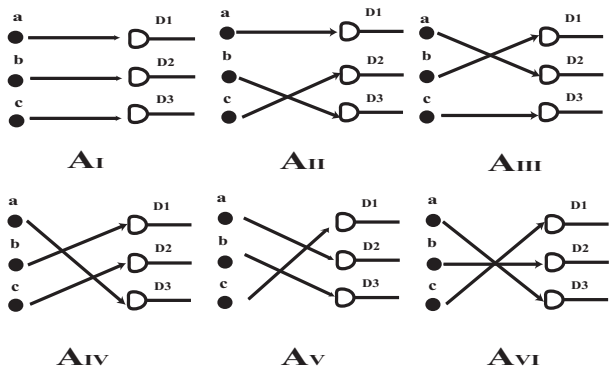}
\caption{Three independent photons $a, b, c$ have six alternative
ways of triggering a joint-detection event between detectors $D_1$,
$D_2$, and $D_3$.  At equal distances from the source, the
probability of observing a three-photon joint-detection event at
$(t_1, t_2, t_3)$ is determined by the superposition of the six
three-photon amplitudes.  At $t_1 =t_2 = t_3$ six amplitudes
superpose constructively. $G^{(3)}(t_1, t_2, t_3)$ achieves its
maximum value by summing over these constructive interferences.
}\label{3photon}
\end{figure}

Assuming equal distances between the source and the photodetectors,
i.e., $z_1 = z_2 = z_3$, the joint detection counting rate of
$D_1$, $D_2$, and $D_3$ is calculated as \cite{Glauber}
\begin{align}\label{amplitude}
& \ \ \ \ G^{(3)}(t_1, t_2, t_3) \nonumber \\
&\propto \int d\omega \, d\omega' \, d\omega''
\big{|}f(\omega)\big{|}^2 \big{|}f(\omega')\big{|}^2 \big{|}f(\omega'')\big{|}^2
\nonumber \\
&\Big{|} \frac{1}{\sqrt{6}} \big{[} g(\omega, t_1)g(\omega', t_2)
g(\omega'', t_3) + g(\omega, t_1)g(\omega'', t_2) g(\omega', t_3)
\nonumber \\
& + g(\omega', t_1)g(\omega, t_2) g(\omega'', t_3) +
g(\omega', t_1)g(\omega'', t_2) g(\omega, t_3) \nonumber \\
& + g(\omega'', t_1)g(\omega, t_2) g(\omega', t_3) + g(\omega'',
t_1)g(\omega', t_2) g(\omega, t_3) \big{]} \Big{|}^2 \nonumber \\
& = \int d\omega \, d\omega' \, d\omega''
\big{|}f(\omega)\big{|}^2 \big{|}f(\omega')\big{|}^2 \big{|}f(\omega'')\big{|}^2
\nonumber \\
& \ \ \times \Big{|} \frac{1}{\sqrt{3!}} \big{[} \sum_{\omega, \omega', \omega''}
g(\omega, t_1)g(\omega', t_2) g(\omega'', t_3) \big{]} \Big{|}^2.
\end{align}
Eq.~(\ref{amplitude}) is the key equation to understand the
three-photon interference nature of the nontrivial third-order
correlation. We notice that the probability amplitude has the
similar form of the symmetrized wavefunction of three identical
particles. The six terms of superposition in Eq.~(\ref{amplitude})
correspond to six different yet indistinguishable alternative ways
for three independent photons to trigger a three-fold
joint-detection event (see Fig.~\ref{3photon}). At $t_1 = t_2 = t_3$
(under the condition of $z_1 = z_2 = z_3$) the six amplitudes are
superposed constructively, and consequently $G^{(3)}(t_1, t_2, t_3)$
achieves its maximum value when summed over these constructive
interferences.  On the other hand, at $t_1 \neq t_2 \neq t_3$, the
six amplitudes are not superposed constructively , the cross terms
equal to zero and consequently $G^{(3)}(t_1, t_2, t_3)$ achieves
lower values.   It is the three-photon interference that caused the
three randomly distributed photons to have 6-times more chance of
being captured at $t_1 = t_2 = t_3$ than that of being captured at
$t_1 \neq t_2 \neq t_3$.

To simplify the calculation, taking the spectral function $f(\omega)$ a
constant within the narrow bandwidth $\Delta \omega$ of the
pseudo-thermal field, the normalized third-order
correlation function $g^{(3)}(t_1, t_2, t_3)$ is approximately
\begin{align}\label{g3-1}
&\ \ \ \  g^{(3)}(t_1, t_2, t_3) \nonumber \\
&= 1 +
\textrm{sinc}^2[\frac{\Delta \omega (t_1 - t_2)}{2}]  \nonumber \\
&+ \textrm{sinc}^2[\frac{\Delta \omega (t_2 - t_3)}{2}] +
\textrm{sinc}^2[\frac{\Delta \omega (t_3 - t_1)}{2}] \\ \nonumber &
+ 2 \,\textrm{sinc}[\frac{\Delta \omega (t_1 - t_2)}{2}]
\textrm{sinc}[\frac{\Delta \omega (t_2 - t_3)}{2}]
\textrm{sinc}[\frac{\Delta \omega (t_3 - t_1)}{2}].
\end{align}
It is easy to see that when $t_1 = t_2 = t_3$, $g^{(3)}(t_1, t_2,
t_3) = 6$,  the third-order correlation function achieves a maximum
contrast of 6 to 1 (visibility $\sim$$71\%$ ). Apparently, the
randomly radiated independent thermal photons seem to have six times
greater chance of being bunched in triples when we measure them
simultaneously than that if we measure when $t_1 \neq t_2 \neq t_3$.

In fact, the quantum theory of light predicts that the randomly
radiated independent ``thermal photons" will have $N!$ times
greater chance of achieving ``$N$-photon bunching" in an $N$-fold
joint-detection of $N$ individual photodetectors.  For large
 $N$, the contrast of the $N$th-order correlation may
achieve $\sim$$100\%$. The calculation is similar to that of the
three-photon \cite{Liu}
\begin{align}\label{GN-0}
& \ \ \ \ G^{(N)}(t_1, ... , t_N) \nonumber \\
&\propto \int d\omega \ ... \ \int d\omega''{}^{...}{}'
\big{|}f(\omega)\big{|}^2 \ ... \ \big{|}f(\omega''{}^{...}{}')\big{|}^2
\nonumber \\
& \ \ \ \ \times \Big{|} \frac{1}{\sqrt{N!}} \big{[} \sum_{\omega ... \omega''{}^{...}{}'}
g(\omega, t_1) ... g(\omega''''{}^{...}{}', t_N) \big{]} \Big{|}^2.
\end{align}
Eq.~(\ref{GN-0}) indicates that the $N$th-order correlation of
thermal radiation is the result of $N$-photon interference.  It is
easy to see that if all of the $N$-photon amplitudes are
superposed constructively at a certain experimental condition, the
thermal radiation will have $N!$ times greater chance to be
measured under that experimental condition.

Although we have observed the nontrivial three-photon correlation
with a contrast much higher than 2 to 1, and expect to have
$N$-photon correlation with a contrast of $N!$ to 1, the quantum
theory does not prevent having a constant counting rate for each of
the photodetectors $D_1$, $D_2$, and $D_3$, respectively.  In fact,
the counting rate of $D_1$, $D_2$, and $D_3$ was monitored during
the measurement and found to be constant, indicating a temporal
randomly emitted and distributed stochastic emission process in the
thermal source. According to the Glauber photo-detection theory, the
counting rate of a photodetector is proportional to the first-order
self-correlation function $G^{(1)}(\mathbf{r},t)$, which measures
the probability of observing a photo-detection event at a space-time
coordinate $(\mathbf{r}, t)$.  For the experimental setup of
Fig.~\ref{scheme},

\begin{equation*}\label{1st}
G^{(1)}(\mathbf{r},t) = \langle \,
E^{(-)}(\mathbf{r},t)E^{(+)}(\mathbf{r},t)\, \rangle = \textrm{constant}.
\end{equation*}

In conclusion, we have observed the nontrivial 3-photon correlation
of pseudo-thermal light.  The photons are always randomly radiated
from the thermal source. This experiment shows a nontrivial
third-order temporal correlation of chaotic-thermal light in which
the randomly radiated thermal light is observed to have a 6-times
greater chance of being captured by three individual photodetectors
simultaneously than that of being captured by the same detectors at
different times (separated by the coherence time of light field).The
nontrivial $N$th-order correlation of thermal light can be
interpreted as the result of multi-photon interference, involving
the superposition of multi-photon amplitudes, a nonclassical entity
corresponding to different yet indistinguishable alternative ways of
triggering a joint-detection event between multiple detectors.

The authors wish to thank M.H. Rubin, J.P. Simon, S. Karmakar,
Z.D. Xie and H. Chen for helpful discussions.  This research was
partially supported by the AFOSR and ARO-MURI program.


\vspace{-3mm}

\begin{thebibliography}{}

\bibitem{HBT} R. Hanbury Brown and R. Q. Twiss, Nature (London) \textbf{177},
27 (1956); \textbf{178}, 1046 (1956); R. Hanbury Brown,
\emph{Intensity Interferometer} (Taylor \& Francis, London, 1974).

\bibitem{Gpaper} G. Scarcelli, V. Berardi and Y. H. Shih, Phys. Rev. Lett., \textbf{96} 063602
(2006).




\bibitem{bunching} L. Mandel and E. Wolf, \textit{Optical coherence and quantum optics} (Cambridge University Press, 1995);
M. O. Scully and M. S. Zubairy, \textit{Quantum Optics} (Cambridge
University Press, 1997);R. Loudon, \textit{The Quantum Theory of
Light} (Oxford University Press, 2000);

\bibitem{Liu} J. B. Liu and Y. H. Shih Phys. Rev. A, \textbf{79} 023819
(2009).

\bibitem{highorder} Y.
F. Bai and S. S. Han, Phys. Rev. A. \textbf{76}, 043828 (2007); D.
Z. Cao, J. Xiong,  S. H. Zhang, L. F. Lin and K. G. Wang,  Appl.
Phys. Lett. \textbf{92}, 201102(2008);  X. H. Chen, I. N. Agafonov,
K. H. Luo, Q. Liu, R. Xian, M. V. Chekhova, and L. A. Wu.
arXiv:0902.3713v1[quantum-ph] (2009)

\bibitem{goodman} J. W. Goodman, \textit{Statistical Optics} (John Wiley \& Sons, Inc., New York, 1985).







\bibitem{klyshko} D. N. Klyshko, \textit{Photons and Nonlinear
Optics} (Gordon and Breach, New York, 1988).

\bibitem{correlationtime} J. Churnside, J. Opt. Soc. Am.
\textbf{72}, 1464 (1982).

\bibitem{Glauber} R. J. Glauber, Phys. Rev., \textbf{84} 10
(1963); R. J. Glauber, Phys. Rev., \textbf{2529} 130 (1963).



\end{thebibliography}
\end{document}